\documentclass{mem}
\usepackage{natbib}\usepackage{txfonts}\usepackage{balance}
\usepackage{flushend}
\usepackage{verbatim}
\usepackage{graphicx}
\usepackage[a4paper,breaklinks,pdftex]{hyperref}
\idline{75}{282}
\begin{document}
\def\teff{$T\rm_{eff }$}
\def\kms{$\mathrm {km s}^{-1}$}

\title{
Margherita Hack's Astrophysics \\at the Asiago Astrophysical Observatory
}

\subtitle{}

\author{
G. \,Umbriaco\inst{1,2,3,5}
\and A. \, Bianchini\inst{1}
\and A. \, Pizzella\inst{1,3}
\and A. \, Siviero\inst{1}
\and G. \, Coran\inst{3}
\and \\P. \, Ochner\inst{1,3}
\and M. \, Montalto\inst{4}
\and M. \, Bazzicalupo\inst{1,3}
\and P. \, Colasanto\inst{1,3}
          }
\institute{
Università degli Studi di Padova
Dipartimento di Fisica e Astronomia "Galileo Galilei"
vicolo dell'Osservatorio 3, I-35122 Padova, Italy
\and
Istituto Nazionale di Astrofisica --
Osservatorio Astronomico di Padova, vicolo dell'Osservatorio 5,
I-35122 Padova, Italy
\and
Università degli Studi di Padova
Museo degli Strumenti dell'Astronomia
via dell'Osservatorio 8, I-36012 Asiago (VI), Italy
\and
Istituto Nazionale di Astrofisica --
Osservatorio Astrofisico di Catania, Via Santa Sofia 78, 
I-95123 Catania, Italy
\and
Dipartimento di Fisica e Astronomia "Augusto Righi" - Alma
Mater Studiorum Università di Bologna, via Piero Gobetti 93/2, 40129
Bologna, Italy
\\
Author contact: \email{gabriele.umbriaco@unipd.it}
}
\authorrunning{Umbriaco}

\titlerunning{Margherita Hack's Astrophysics at the Asiago Observatory}

\date{Received: Day Month Year; Accepted: Day Month Year}

\abstract{

Margherita Hack observed with the Galileo telescope at the Asiago Astrophysical Observatory from September 1951 to March 1954. Using the spectroscopic facilities of the observatory, Margherita contributed to the stellar study of novae, symbiotic stars, and peculiar stars. In the 80th anniversary of the Asiago Astrophysical Observatory and the 100th year of Margherita, we found her observations in the Asiago Photographic Plate Archive and we remade some Margherita's observations with the current capabilities of Asiago telescopes. We present here a summary of the early studies carried out by Margherita with the Galileo telescope. We underline the importance of maintaining a well-organised plate archive to allow historical and scientific studies.

}
\maketitle{}

\section{Introduction}
\begin{figure}
\resizebox{\hsize}{!}{\includegraphics[clip=true]{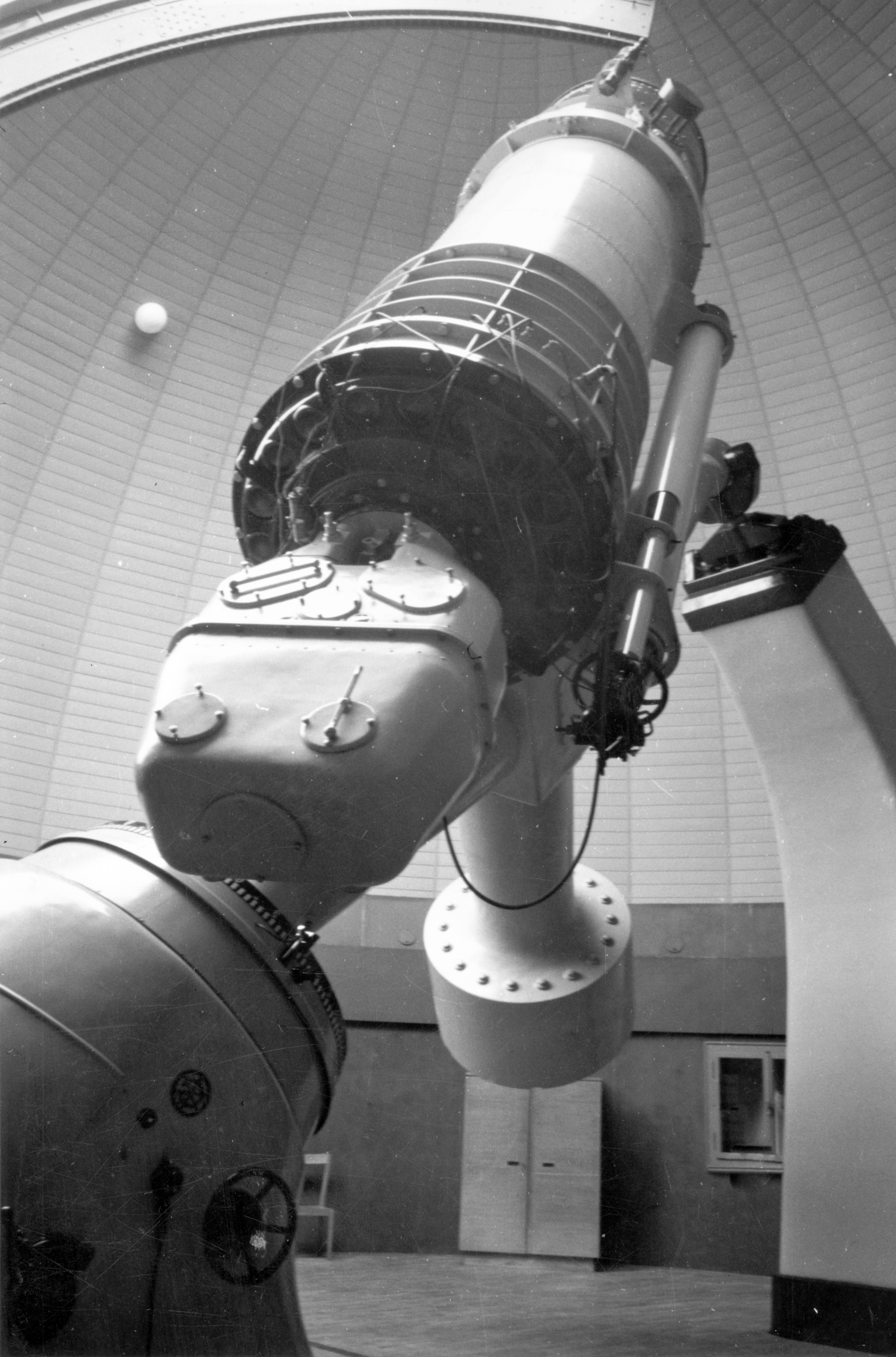}}
\caption{
\footnotesize
The Asiago Observatory with the Prism Spectrograph "A" mounted to the Cassegrain focus (1946).  Credits: \href{https://phaidra.cab.unipd.it/detail/o:328466?mycoll=o:330194}{Phaidra Digital Collection University of Padova}}
\label{t122}
\end{figure}

The aim of this article is to show Margherita Hack's early astrophysical observations at the Asiago Astrophysical Observatory, by using the Asiago Prism Spectrograph "A", see Fig.\ref{t122}, capable of high resolution in a wide range of the spectrum, one of the first spectrograph coupled with a large telescope in Italy since 1946. 
We will show how information on Margherita's observations at Asiago was found in the Asiago Photographic Plate Archive, emphasising the important scientific and historical role of astronomical archives. Finally, we will show how bright Margherita was at guessing some spectral features on the photographic plates, which are evident today with the instrumental improvement adopted by the Galileo telescope.
Margherita investigated the astrophysics of peculiar stars including: stars located in the main sequence like hot blue, blue supergiants and white hypergiants with spectral type O, B, Be, A. She also studied binary systems composed of different kinds of stars: hot and massive stars, eclipsing binaries, subgiants, eclipsing Algol-type binaries, binary systems composed of two subgiant stars belonging to the A-type or stellar systems composed of three stars, or binary systems composed of a white subgiant and less massive main-sequence dwarf star.
These observations helped Margherita to conduct scientific researches that covered a wide range of topics. Interpreting the spectroscopic characteristics of stars became her main area of expertise, involving the study of the chemical composition of stars and of their surface temperature and gravity. Margherita improved the spectroscopic observations by comparing the capabilities of the Asiago Observatory spectrograph with the objective prism applied to the Schmidt telescope of the Arcetri Observatory and other spectrograph.

\begin{figure}
\resizebox{\hsize}{!}{\includegraphics[clip=true]{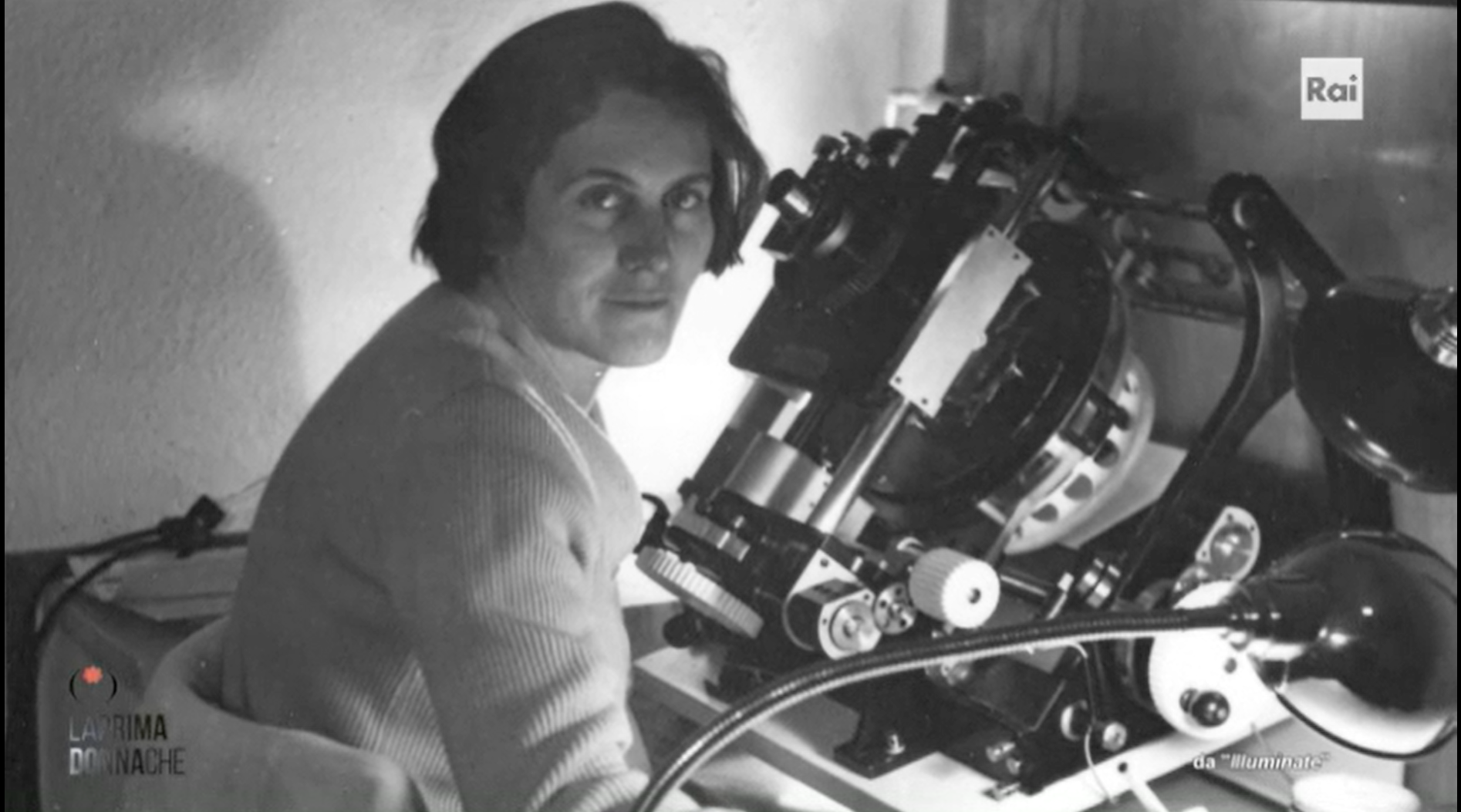}}
\caption{\footnotesize
The image shows Margherita working at the Mioni comparator also present in the Museum of the Asiago Observatory. Credits: \href{https://www.raiplay.it/video/2021/05/la-prima-donna-che-margherita-hack-3c508aa0-73a8-4977-9238-454474877305.html}{RAI-Documentari}.
}
\label{hack.mioni}
\end{figure}

Margherita observed at the 1.22 m Galileo Telescope from 9 September 1951 to 31 March 1954, collecting over 180 spectra with the Prism Spectrograph 'A', reached a cumulative exposure time exceeding 100000 seconds on 15 different targets. In Fig.\ref{hack.mioni} Margherita is working with the Mioni plate analyser; a copy of this instrument is currently available in the Museum of Astronomical Instruments of the Asiago Astrophysical Observatory.

Margherita's experience in spectroscopic observations, including those carried out in Asiago, and her interest in the stellar spectroscopy of normal and peculiar stars formed the basis of books \textbf{Stellar Spectroscopy} \cite{1969stsp.book.....H} \cite{1970stsp.book.....H} published in 1969 and 1970 with coauthor Otto Struve. The books became a classic reference text for the study of stellar atmospheres.

\section{Margherita's scientific achievements from observations in Asiago}
For each object observed by Margherita with the spectrograph at Asiago, one or more publications were produced, comparing measurements also obtained with other telescopes. This work gave Margherita scientific information and knowledge that led to a deep understanding of stellar astrophysics, summarised below. 

By taking 16 spectrograms at the Asiago 'Galileo' telescope in September 1951, Hack measured and reduced the astronomical plates to derive the quantitative analysis of the atmosphere of \textbf{55 Cygni} \cite{1953MmSAI..24...31H} and \cite{1953MmArc..68...63H}, using the curve of growth theory. She obtained an average velocity of the atmospheric turbulence of 46 km/sec from the H and He lines. Also she was able to measure a temperature gradient of 55 Cyg w.r.t. $\alpha$ Cyg. Hack figured out this anomaly due to the presence of interstellar reddening or to a peculiar constitution of 55 Cyg. 

Hack made a comparison between the results obtained in the quantitative analysis of star 55 Cygni, spectrum B2, by comparing \cite{1953MmArc..68..125H} \cite{1953MmSAI..24...89H} spectrograms made with an objective prism at Arcetri and a slit-spectrograph at Asiago with dispersion, respectively: 185 $\mathring{A}$/mm and 40 $\mathring{A}$/mm at H$\gamma$. Hack concluded that the results obtained with the objective prism are in fairly good agreement with those obtained with the slit spectrograph for the determination of the abundance of hydrogen and helium atoms and the electronic density.

Hack studied the supergiant star \textbf{A3 Ia 6 Cassiopeiae} from 40 $\mathring{A}$/mm at H$\gamma$ dispersion spectrograms \cite{1954MmSAI..25..481H}. She found that the main features of the spectrum were in the Balmer lines and confirmed the spectral type A5 Ia. Comparison of the observed profiles of H$\gamma$ and H$\delta$ with the theoretical profiles calculated by Verweij led to the best agreement for log(g) = 1 and T = 12600°. She compared the relative abundances for 6 Cas, the Sun and 55 Cyg and deduced that: 1) there is no essential difference of chemical composition; 2) the atmosphere opacity of 6 Cas is similar to that of the Sun, and the opacity of 55 Cyg, B3 Ia, is nearly 40 times greater than 6 Cas.

Using spectrograms taken from the spectrograph at the Asiago Observatory, Hack studied the spectrum of $\zeta$ Tauri \cite{1955MmSAI..26...41H}. She found, using two dispersions of 40 $\mathring{A}$/mm and 15 $\mathring{A}$/mm at H$\gamma$, that the rise of the central depths of the H lines was considerable. She constructed two branches of the growth curve with the Fe II and O II lines, representing the shell and the atmosphere of the star respectively. This led to obtaining a turbulence velocity of 7 km/sec for the shell and 23 km/sec for the star. Hack took several spectrograms from November 1965 to March 1966 using the Asiago prism spectrograph with a dispersion of 40 $\mathring{A}$/mm at H$\gamma$ \cite{1967MmSAI..38..623H}. Subsequent observations revealed that the asymmetry had completely disappeared, while H$\alpha$ and H$\beta$ still exhibited emission wings with R/V $> 1$.

%%% Using spectrograms taken from the spectrograph at the Asiago Observatory, Hack studied the spectrum of \bm{$\zeta$} \textbf{Tauri} \cite{1955MmSAI..26...41H}. She found, using the two dispersions of 40 $\mathring{A}$/mm and 15 $\mathring{A}$/mm at H$\gamma$, that the rise of the central depths of the H lines was considerable. She constructed two branches of the growth curve with the Fe II and O II lines regarding, respectively, the shell and the atmosphere of the star, obtaining a velocity of turbulence of 7 km/sec for the shell and 23 km/sec for the star. Hack took several spectrograms from November 1965 to March 1966 with the Asiago prism spectrograph (dispersion 40 $\mathring{A}$/mm at H$\gamma$) \cite{1967MmSAI..38..623H}. They showed that the asymmetry has completely disappeared and that H$\alpha$ and H5 still had emission wings with R/V $> 1$.

Hack gave the results of the silicon star \textbf{HD 34452} evaluating: a) the most probable spectral type and luminosity class; b) the abundances of H and He following the Unisiild method; c) the other element abundances (C, N, Mg, Si, Fe) and the turbulence velocity (2,6 km/sec) from the curve of growth; d) the exceptional intensity of the Si II and Mg II lines, which were the main peculiarity of the HD 34452 spectrum, consequently giving an excess for the relative abundances of these elements. This excess has been explained by a real difference in the chemical composition or by the presence of a stellar magnetic field \cite{1956MmSAI..27..307C}.

Hack and T. Tamburini compared the results of their spectroscopic observations in March 1954 \cite{1958MmSAI..29..165H} for ${\alpha}^{2}$ Canum Venaticorum (${ \alpha}^{2}$ C Ven) with the results obtained by \cite{1943ApJ....98..361S}, \cite{1955ApJS....1..431B}, and \cite{1939MNRAS.100...94T}. They found good agreement with the variations in intensity of the lines of Si II, Cr II, and Ca II. However, the results were not in agreement with the lines of Fe II, Mg II, and Sr II. Additionally, they compared the spectroscopic data for ${\alpha}^{2}$ C Ven, ${\epsilon}$ Ursae Majoris (AOV p), and ${\alpha}$ Lyrae (AOV), suggesting that the spectral type of ${\alpha}^{2}$ C Ven was likely B8 V. They also identified the main peculiarities of its spectrum, including a large number of weak lines of metals and rare earths, an exceptional intensity of Si II lines, and a weak intensity of the K line.

%%% Hack and T.Tamburini compared the results of their spectroscopic observations of March 1954 \cite{1958MmSAI..29..165H} for \bm{${\alpha}^{2}$} \textbf{C Ven} with the results obtained by \cite{1943ApJ....98..361S}, \cite{1955ApJS....1..431B} and of \cite{1939MNRAS.100...94T}. They found good agreement with the variations in intensity of the lines of Si II, Cr II, and Ca II; while the results disagreed with those of the lines of Fe II, Mg II and Sr II. They compared the spectroscopic data for ${\alpha}^{2}$ C Ven, ${\epsilon}$ U Ma (AOV p) and ${\alpha}$ Lyrae (AOV) showing that probably the spectral type of ${\alpha}^{2}$ C Ven was B8 V. They found that the main peculiarities of its spectrum are the large number of weak lines of metals and rare earths, the exceptional intensity of Si II lines, and the weak intensity of the K line.

Hack and Aydin studied the variation of the spectrum as a function of time of \textbf{Symbiotic Star CH Cyg} from observations made on Haute Provence Observatory, some spectra obtained at Asiago Copernico telescope by Iijima between January 1990 and June 1991 and ultraviolet observations by IUE \cite{1992Ap&SS.194..215H}. They found an increase in activity in this binary system between July and December 1990. This increase was confirmed in the UV range by IUE observations. A more general discussion of Symbiotic Variables was provided \cite{1993NASSP.507..663V} and they discussed a number of individual objects for which the amount of observational data was large enough to draw a fairly complete picture of their behaviour. Since this was a useful occasion to describe different diagnostic techniques, they illustrated the steps toward a possible empirical model.

Hack was also attracted by the spectacular outbursts of \textbf{Classical novae and Recurrent novae}, so she decided to review the phenomenology of their complex phenomenon in a special NASA publication. Due to the very complicated individual behaviour of each nova, it was chosen to review the observations of a few well-observed classical \cite{1993NASSP.507..413H} and recurrent novae of different brightness and speed classes \cite{1993NASSP.507..511H} from observations of different observatories, including Asiago.

\vspace{-1mm}

\section{Asiago Photographic Plate Archive}
\begin{figure}
\resizebox{\hsize}{!}{\includegraphics[clip=true]{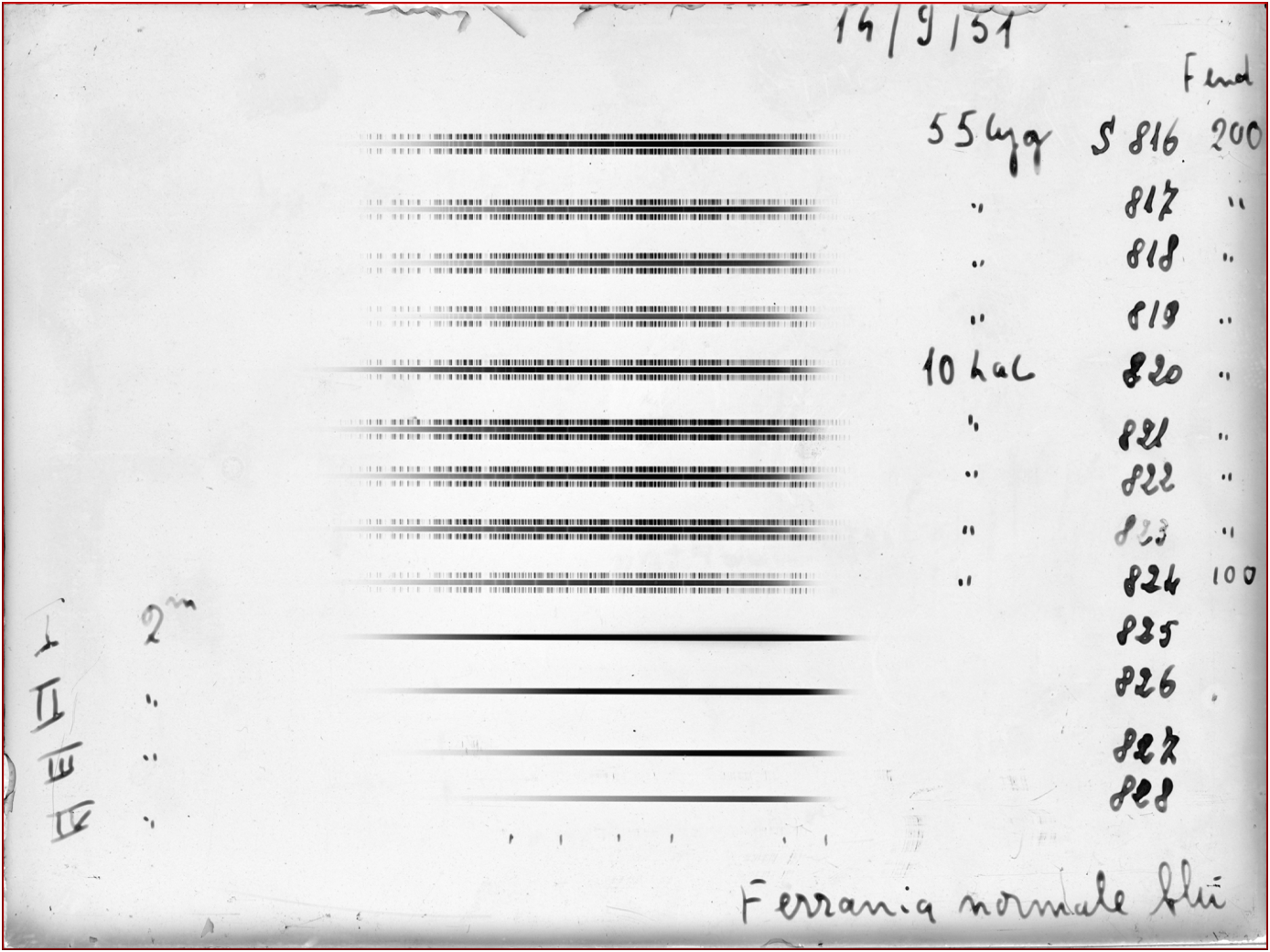}}
\caption{
\footnotesize
The photographic plate of Margherita preserved in Asiago.
}
\label{first-plate}
\end{figure}

The Plate Archive in the Asiago Astrophysical Observatory preserves 83000 plates collected by the telescopes at the University of Padova and Istituto Nazionale di Astrofisica - INAF since the beginning of the observations, as listed in table \ref{abun}. The archive also preserves the observation notebooks and logbooks. Before the digital age, plates were measured using optical comparators and densitometers, today digital scanners are used, e.g. we used Epson Expression 10000XL and 1640XL professional scanners. 
Looking for Margherita's original plates, we discovered in the \href{https://doi.org/10.5281/zenodo.7517003}{log book of observation} 141 observations made in Asiago by Hack, but only one plate was left in the Asiago Photographic Plate Archive. As was often the practise, astronomers took the plates of their observations with them to continue studying them in their observatories. Our experience shows that the plates kept by the astronomers at our institute are returned to the archive after retirement. A note on their relocation is usually kept in the archive, whereas not in the case of Margherita's plates. Margherita's original plates will probably be in the observatories where Margherita worked or in some private place. In these cases we recommend returning the plates to our observatory for long-term preservation, and we give the possibility of delivering a digitised copy made on site; refer to the author of this paper. In Fig.\ref{first-plate} the scanned version of the spectra of 55 Cygni and 10 Lacertae taken by Hack during 14 and 15 September 1951 at the Asiago Observatory.

\begin{table}[h]
\caption{Asiago Plates Archive}
\vspace{-4mm}
\label{abun}
\begin{center}
\begin{tabular}{lccccccc}
\hline
\\
Telescope & Number of plates\\
\hline
\\
Galileo 1.22m    &$ 36923 $ \\
Copernico 1.82m  &$ 8171 $  \\
Schmidt 50/40    &$ 17816 $ \\
Schmidt 92/67    &$ 20417 $ \\
\\
\hline
\end{tabular}
\end{center}
\vspace{-4mm}
\end{table}

\section{$\zeta$ Tau re-observed in 2022}
Margherita used the Galileo telescope equipped with the spectrograph "A", which suffers from differential dispersion due by using two prisms, from 13$\mathring{A}$/mm in blue to 150$\mathring{A}$/mm in red. In spite of this, she was able to recognise spectral characteristics that are only evident today, thanks to the use of grating spectrographs equipped with digital sensors. 
In recent years, the Galileo telescope was upgraded with a grating spectrograph, the Boller\&Chivens by INAF \cite{2012BaltA..21....1R} \cite{2014CoSka..43..362C}. The B\&C spectrograph was coupled with a high quantum efficiency CCD ANDOR iDus DU440A with low noise and dark current. 
We propose here a comparison between the Hack observation of $\zeta$ Tau made in 1954 and those in 2022 by using the same Galileo telescope and also the Copernico telescope. In these observations, we used 1200 grooves/mm blazed at 6825$\mathring{A}$ capable of 44$\mathring{A}$/mm. For high-resolution spectra instead, we used the Echelle spectrograph mounted at the 1.82m Copernico telescope operated by INAF, in Asiago too. The Echelle spectrograph is equipped with an Andor iKon DW436-BV, equipped with a 300 grooves/mm grating at $-8.5^{\circ}$ capable of a resolution of 8$\mathring{A}$/mm.

\begin{figure}
\resizebox{\hsize}{!}{\includegraphics[clip=true]{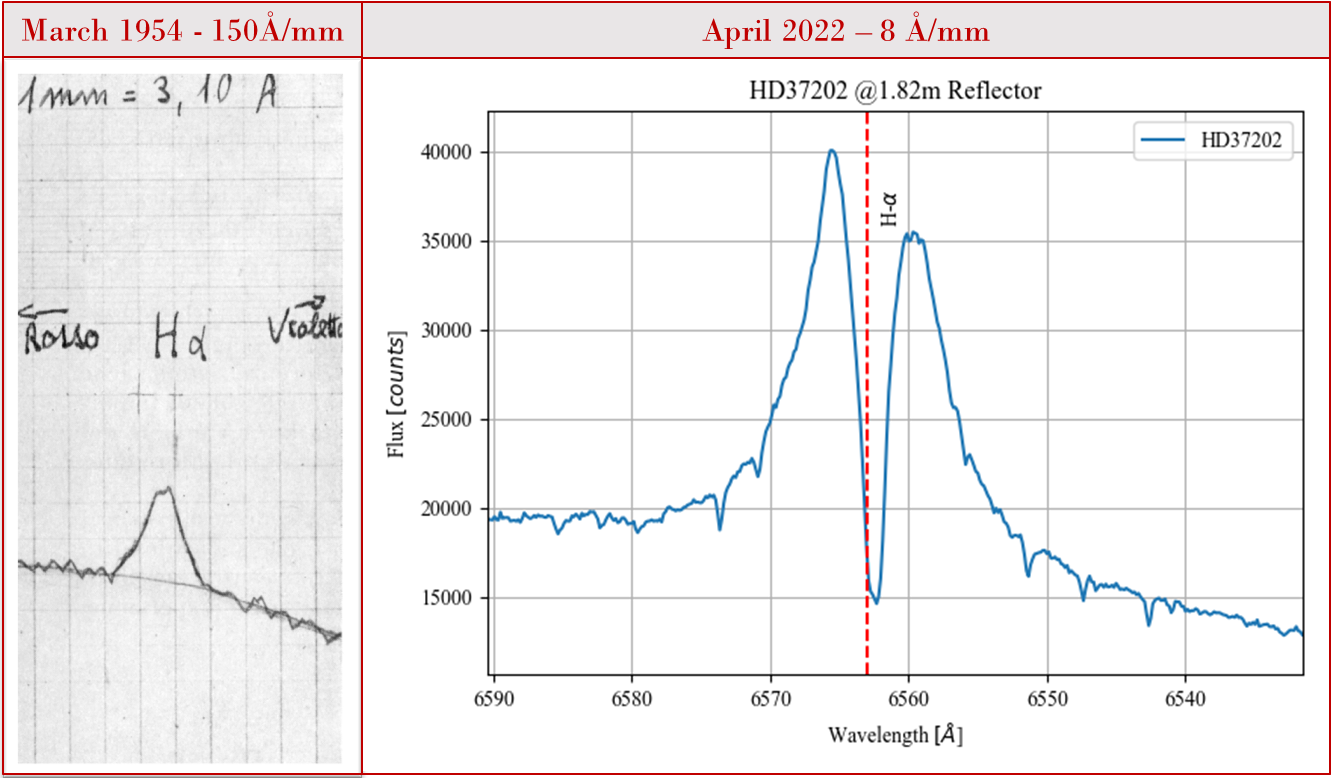}}
\caption{
\footnotesize
The H$\alpha$ profile of $\zeta$ Tau at Galileo (1954, left) and at Copernico (2022, right).
}
\label{HD37202-Halpha}
\end{figure}

\begin{figure}
\resizebox{\hsize}{!}{\includegraphics[clip=true]{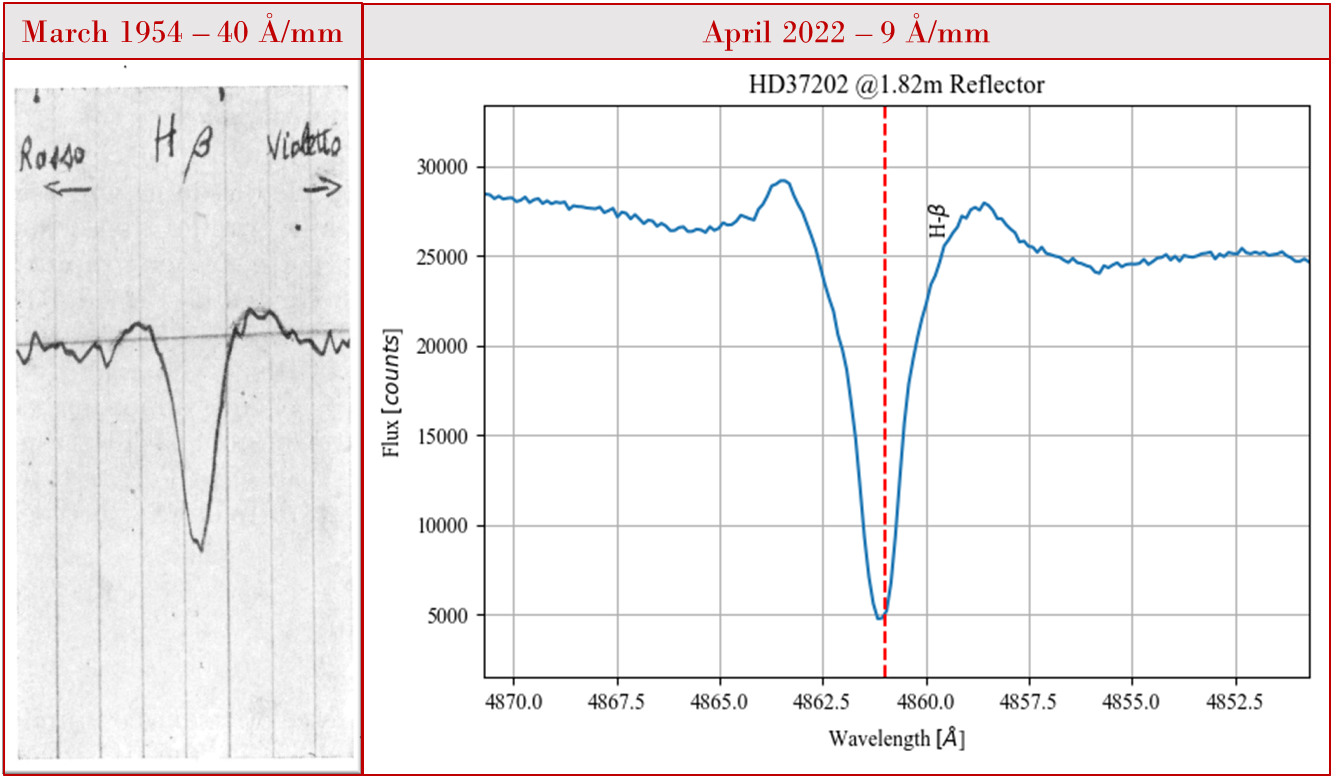}}
\caption{
\footnotesize
The H$\beta$ profile of $\zeta$ Tau at Galileo (1954, left) and Copernico (2022, right).
}
\label{HD37202-Hbeta}
\end{figure}

\begin{figure}
\resizebox{\hsize}{!}{\includegraphics[clip=true]{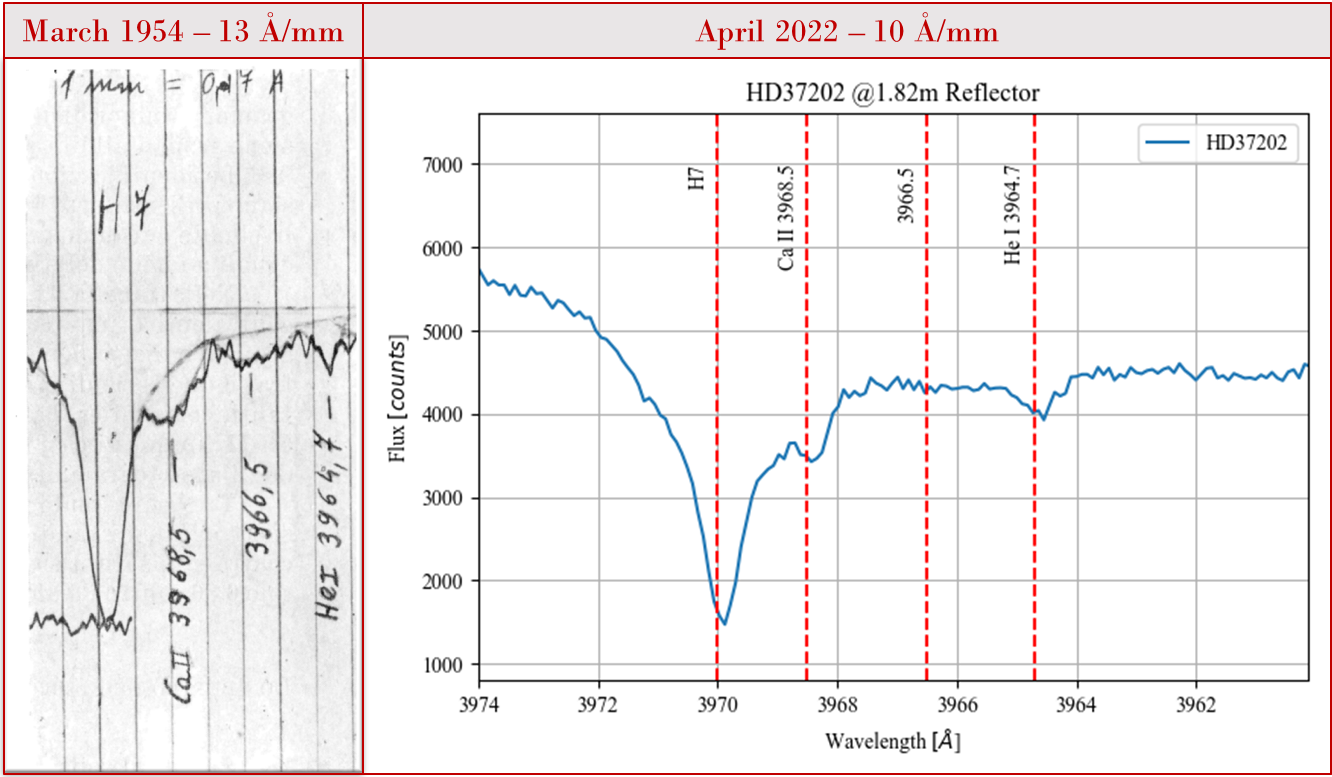}}
\caption{
\footnotesize
The H$7$ profile of $\zeta$ Tau at Galileo (1954, left) and Copernico (2022, right).
}
\label{HD37202-H7}
\end{figure}

The observation with the Galileo telescope was carried out on March 6, observing some targets, including reference spectrophotometric stars: HR1544, HD37202, V725 Tau, Eta Aur, Epsilon UMa, alf02CVn; at the Copernico telescope on April 19 the targets were HD37202 and HR3454. 
Here, we report in Fig.\ref{HD37202-Halpha}, Fig.\ref{HD37202-Hbeta} \& Fig.\ref{HD37202-H7} the observation of $\zeta$ Tau (HR 1910, HD 37202, HIP 26451), a bright Be star with an extensive observational history that spans photometric, spectroscopic, polarimetric, and interferometric techniques. The spectrum of $\zeta$ Tau shows the standard double-peaked H$\alpha$ profile, indicative of a disk in Keplerian rotation. The relative heights of the blue- and red-shifted peaks of the emission lines (V/R ratios) show cyclic variability measured on the order of 1429 days. The spectra of $\zeta$ Tau also occasionally show more complex triple-peaked and shell profiles. Current models suggest that the line profile variations in $\zeta$ Tau and other Be stars can be explained by global one-armed oscillation models \cite{2010AJ....140.1838S}. 

\begin{figure}
\resizebox{\hsize}{!}{\includegraphics[clip=true]{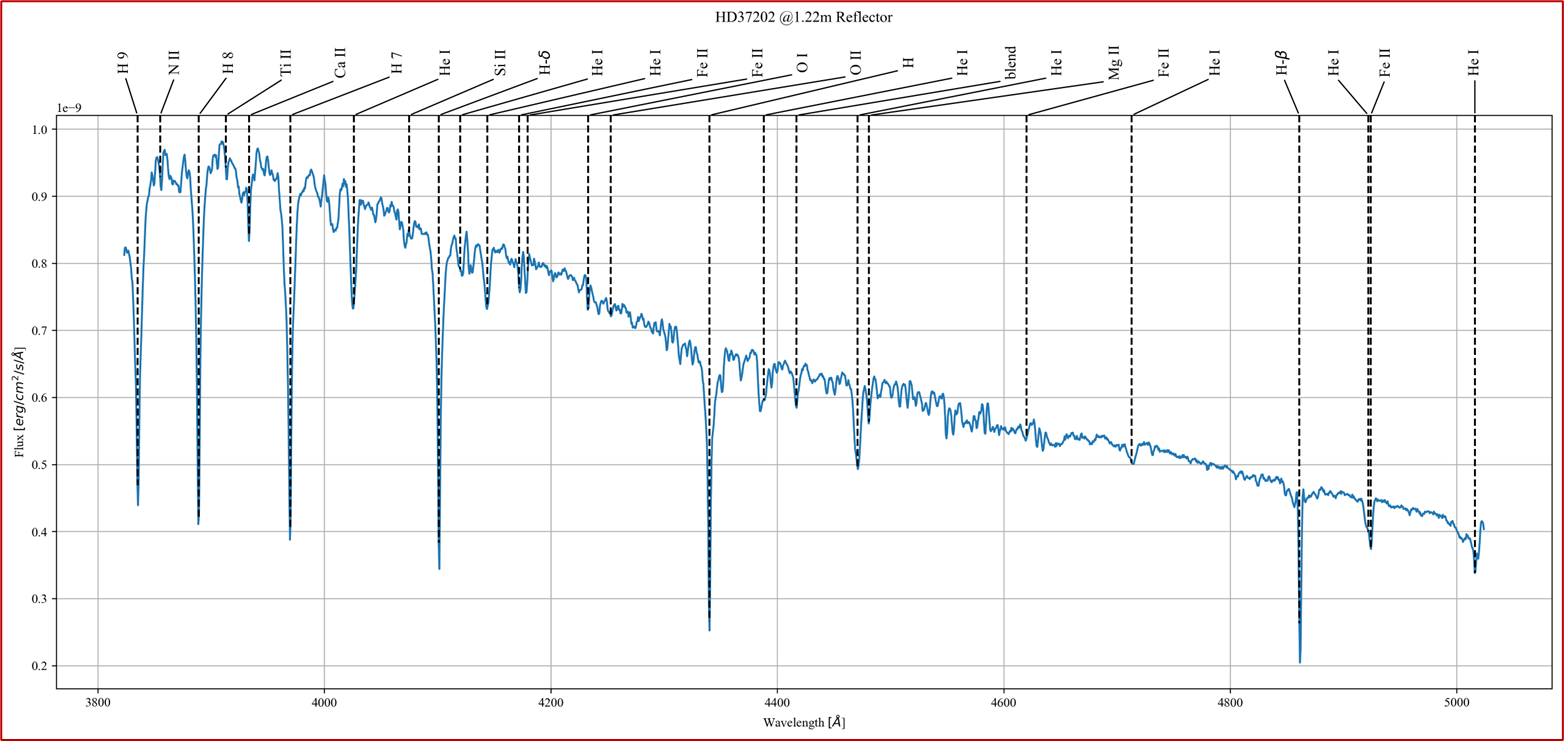}}
\caption{
\footnotesize
The spectrum profile of $\zeta$ Tau at Galileo telescope (2022) with overimposed the lines identified by Hack in the observations of 1954.
}
\label{HD37202-T122}
\end{figure}

In Fig.\ref{HD37202-Halpha} Hack had only 150 $\mathring{A}$/mm at 6536 $\mathring{A}$ despite the low dispersion of the prism spectrograph in the red part of the spectrum, she recognised that the broadened tip of the hydrogen spectrum could be associated with weak absorption. The absorption is evident in the spectrum obtained from the Echelle spectrograph of the 182 cm Copernicus telescope (right). In Fig.\ref{HD37202-Hbeta} Hack shows that for H$\beta$ the emitting wings have a more intense blue component than the red one. In 2022, using the Echelle spectrograph of the 182 cm Copernicus telescope, this feature appears reversed on the star, with the blue wings less intense than the red ones. In Fig.\ref{HD37202-H7} Hack recognises the non-rotational profile characteristic of the stellar shell, and the Echelle spectrogram of the Copernicus telescope confirms these features also in 2022. 
The spectrum in Fig.\ref{HD37202-T122} shows the $\zeta$ Tau spectrum obtained with the 1.20m 'Galileo' telescope with overimposed lines identified by Hack in the observations of 1954. The equipments and measurement techniques used by Margherita Hack were nor as precise and straightforward as today's digital ones, which emphasises even more how much work was behind the scientific results she obtained some 70 years ago.

\section{Conclusions}
In this paper, we reviewed the observations of Margherita Hack carried out in the 1951-54 period, at the Asiago Astrophysical Observatory, and revisited in 2022. We underline the importance of keeping a well-organised plate archive, useful both for historical studies on astronomical targets and for the cultural heritage of important astronomers, such as Margherita was.

\begin{acknowledgements}
 A particular acknowledgement to volunteers of alternative civil service of the Museum of Astronomical Instrumentation who worked in the Asiago Astrophysical Observatory on the conservation of the Astronomical Plates Archive.
\end{acknowledgements}

\small
\bibliographystyle{aa}
\bibliography{biblio}

\end{document}